\documentclass[pra,aps,showpacs,superscriptaddress]{revtex4}
\usepackage{bm}
\usepackage{graphicx}
\usepackage{amsbsy}
\usepackage{amsmath}
\usepackage{amsfonts}

\newcommand{\bra}[1]{\langle{#1}|}
\newcommand{\ket}[1]{|{#1}\rangle}

\newcommand{\nn}{\nonumber}

\begin{document}

\title{Equivalence between two-mode spin squeezed states \\
	and pure entangled states with equal spin}

\author{Dominic W.\ Berry} 
\affiliation{Department of Physics, The University of Queensland, St. Lucia,
	Queensland 4072, Australia}
\affiliation{Centre of Excellence for Quantum Computer Technology,
	Macquarie University, Sydney, New South Wales 2109, Australia}
\affiliation{Institute for Quantum Information Science, University of Calgary,
	Alberta T2N 1N4, Canada}
\author{Barry C.\ Sanders} 
\affiliation{Centre of Excellence for Quantum Computer Technology,
	Macquarie University, Sydney, New South Wales 2109, Australia}
\affiliation{Institute for Quantum Information Science, University of Calgary,
	Alberta T2N 1N4, Canada}

\begin{abstract}

We prove that a pure entangled state of two subsystems with equal spin
is equivalent to a two-mode spin-squeezed state under local operations except
for a set of bipartite states with measure zero, and we provide a counterexample
to the generalization of this result to two subsystems of unequal spin.

\end{abstract}
\date{\today}


\maketitle

\section{Introduction}
The uncertainty principle is a consequence of the noncommutativity of
complementary variables. Specifically, two complementary operators $A$ and $B$
have spreads $\Delta A$ and $\Delta B$, given by roots-of-variances
$\sqrt{V(A)}$ and $\sqrt{V(B)}$, respectively, that satisfy the inequality
$\Delta A \Delta B \ge |\langle C \rangle|/2$ where $iC=[A,B]$. Often natural
units can be employed such that $A$ and $B$ are dimensionally equivalent and
$\Delta A \ge  \sqrt{|\langle C \rangle|}$ and $\Delta B \ge \sqrt{|\langle C
\rangle|}$ for a typical state. Under these conditions, $\sqrt{|\langle C
\rangle|}$ represents a fundamental noise limit, which is known as the
\emph{standard quantum limit}. This standard quantum limit is especially
important in quantum metrology. One example is optical interferometry, for
which semiclassical input states (which can be expressed as a mixture of
Glauber-Sudarshan coherent states) have, at best, vacuum fluctuations; in this
case $A$ and~$B$ are the canonical harmonic oscillator operators, and each of
$\Delta A$ and $\Delta B$ exceed $\tfrac{1}{2}$. Spin systems, whose dynamical
operators $J_x$, $J_y$, and $J_z$, satisfy $iJ_z=[J_x,J_y]$, are another
example: the standard quantum limit for $\Delta J_x$ and $\Delta J_y$ is
$\sqrt{|\langle J_z\rangle|}$. An objective of quantum metrology is to prepare
states whose noise level is less than these standard quantum limits; such states
are generically known as `squeezed' because of the reduction, or squeezing, of
the fluctuations below the standard quantum limit.

Spin squeezing, for which the uncertainty in one spin component is reduced
below the standard quantum limit, has been studied extensively~\cite{all,Win94,
Sor01a,Devi03a} and is especially important for applications to high-resolution
spectroscopy~\cite{Win94}. Furthermore, spin squeezing has become important
because spin squeezing implies entanglement within the spin
system~\cite{Sor01a}. For two spin-1/2 systems, pure entangled states are
equivalent to spin-squeezed states under local unitary
transformations~\cite{Devi03a}. For bipartite spin systems, pure two-mode
spin-squeezed (TMSS) states~\cite{Kuz00,Juls01,Ber02a,Ber02b,Ray03}, which
exhibit strong correlations between spin components of the two subsystems, have
been shown to be entangled~\cite{Ber02b,Ray03}; that is, two-mode spin
squeezing is a sufficient condition for entanglement.

Thus we know that two-mode spin squeezing implies entanglement for two-mode 
spin systems. The open question is whether entanglement is equivalent to
two-mode spin squeezing up to local unitary transformations. Here we show that,
for pure states of two spin systems of equal dimension, two-mode spin squeezing
after application of local unitaries is a \emph{necessary} condition for
entanglement, except for a set of bipartite pure states of measure zero. Thus
two-mode spin squeezing can be considered to be approximately equivalent to
entanglement, in the sense that the exceptions are of measure zero.
Furthermore we show that two-mode spin squeezing is not equivalent to
entanglement for (a)~mixed states, (b)~two spin systems of unequal dimension, or
(c)~the restriction of the local unitary operations to rotations. For these
cases, the set of exceptions has nonzero measure.

\section{Equivalence}
We apply the superscript $(1)$ or $(2)$ to the spin operators~$J_k$ to indicate
the subsystem. We take these subsystems to each be of fixed total spin; that is,
we do not consider superpositions of different total spins. In this section
we take both subsystems to have the same spin, $j$. Sums and differences of the
spin operators are denoted by $J_k^{(\pm)} = J_k^{(1)} \pm J_k^{(2)}$, and the
usual criterion for two-mode spin squeezing may then be expressed as
\begin{equation}
\label{eq:crit}
	V(J_y^{(+)}) + V(J_x^{(-)}) < \langle J_z^{(+)} \rangle .
\end{equation}
Any pure entangled state may be expressed, via a Schmidt decomposition, as
$\ket{\Psi} = \sum_{m=-j}^j \psi_m \ket{\varphi_m}\ket{\chi_m}$,
where $\psi_m$ are the Schmidt coefficients, and $\ket{\varphi_m}$ and
$\ket{\chi_m}$ are orthonormal bases for the two spin systems. The Schmidt
coefficients are nonnegative real numbers, and we label them such that they are
in nondescending order. 

Using local unitary operations, we map the bases $\ket{\varphi_m}$ and
$\ket{\chi_m}$ to the bases of $J_z$ eigenstates, thus obtaining
\begin{equation}
\label{eq:psi}
	\ket{\psi} = \sum_{m=-j}^j \psi_m \ket{m,m}_z
\end{equation}
with $|m_1,m_2\rangle_z \equiv |m_1\rangle_z|m_2\rangle_z$.
The state~(\ref{eq:psi}) satisfies certain symmetry conditions. The first is
that the expectation values of the $x$- and $y$-spin components are zero.
To see this, note that $J_z^{(-)}\ket{\psi}=0$ so
\begin{equation}
\label{same1}
\langle J_{x,y}^{(+,-)} \rangle = - \langle e^{-i\pi J_z^{(1)}}
J_{x,y}^{(+,-)} e^{i\pi J_z^{(1)}} \rangle  = -\langle J_{x,y}^{(+,-)} \rangle
\end{equation}
where we have used the subscripts $x,y$ and superscripts $(+,-)$ to indicate
that the same derivation holds for any of these operators. Thus we have 
$\langle J_y^{(+)} \rangle = \langle J_x^{(-)} \rangle = 0$; hence
\begin{equation}
	V(J_y^{(+)}) = \langle (J_y^{(+)})^2\rangle, \quad 
	V(J_x^{(-)}) = \langle(J_x^{(-)})^2\rangle.
\end{equation}
We may show that these variances are equal using a similar method:
\begin{equation}
\label{same2}
\langle (J_y^{(+)})^2\rangle = \langle e^{-i\frac{\pi}2 J_z^{(-)}}(J_y^{(+)})^2
e^{i\frac{\pi}2 J_z^{(-)}}\rangle = \langle (J_x^{(-)})^2 \rangle .
\end{equation}

Therefore, in order to show that the state $\ket{\psi}$ is a TMSS state, it is
sufficient to establish that
\begin{equation}
\label{eq:simpcrit}
	\langle (J_x^{(-)})^2 - J_z^{(+)}/2 \rangle < 0.
\end{equation}
We can show this result by simply evaluating the left-hand side. Expanding 
Eq.~(\ref{eq:simpcrit}) yields
\begin{equation}
\label{eq:three}
\langle (J_x^{(-)})^2 - J_z^{(+)}/2 \rangle = 2\langle (J_x^{(1)})^2 \rangle
-2\langle J_x^{(1)}J_x^{(2)} \rangle - \langle J_z^{(+)}/2 \rangle
\end{equation}
where we have used $\langle (J_x^{(1)})^2 \rangle=\langle(J_x^{(2)})^2\rangle$,
which follows from symmetry. From the calculations in Appendix~A, we obtain
\begin{equation}
	\label{final}
	\langle (J_x^{(-)})^2 - J_z^{(+)}/2 \rangle = \sum_{m=-j}^{j-1}
	(\psi_m-\psi_{m+1})\psi_m [j(j+1)-m(m+1)].
\end{equation}
Because the $\psi_m$ are nonnegative and in nondescending order,
$(\psi_m-\psi_{m+1})\psi_m \le 0$. In addition, if the Schmidt coefficients
take more than one nonzero value, there must be a value of $m$ such that
$\psi_m$ and $\psi_{m+1}$ are not equal and both nonzero, which implies that
$(\psi_m-\psi_{m+1})\psi_m<0$.

Therefore, provided the nonzero Schmidt coefficients of $\ket{\psi}$ are not all
equal, $\langle (J_x^{(-)})^2 - J_z^{(+)}/2 \rangle$ is stricly less than zero,
and $\ket{\psi}$ is a TMSS state. Thus we have shown that all states with
Schmidt coefficients that take more than one nonzero value are equivalent to
TMSS states under local unitary operations.

\section{Case of equal Schmidt coefficients}
For the case where the nonzero Schmidt coefficients of the state are all equal,
the proof given in the previous section does not apply. This case includes
(i)~unentangled states, (ii)~maximally entangled states with \emph{all} 
Schmidt coefficients equal, and (iii)~states with some of the Schmidt
coefficients zero, and the remainder equal. For the third case, if we restrict
to the subspaces for the two subsystems that are spanned by the states in the
Schmidt decomposition, the state is maximally entangled. For case (i), the
states are unentangled, so it is clear that they can not be equivalent to TMSS
states~\cite{Ber02b,Ray03}. Cases (ii) and (iii) comprise a set of bipartite
pure spin states of measure zero so, although these states can be exceptions to
the principle of equivalence between TMSS and entanglement, they are rare in
that the probability for such states is zero when selected according to the
Haar measure.

It is easily seen that, for case (ii), the states are counterexamples to the
principle of equivalence between TMSS and entanglement. For maximally entangled
states, the reduced density operator for each subsystem is the identity, so
$\langle J_z^{(+)} \rangle=0$, and the state clearly cannot be TMSS under any
local unitaries. For case (iii), the state $\ket\psi$ in the form
\eqref{eq:psi} is not TMSS. This does not eliminate the possibility that there
are local unitary operations that bring the states to TMSS form. However,
numerical searches have failed to find such operations.

\section{Extensions to other cases}
The results that we have presented have three requirements; that the spins are
equal, the states are pure, and arbitrary local unitary operations are allowed.
Below we present examples demonstrating that if any of these three requirements
are lifted, by allowing unequal spin, mixed states, or restricting to local
rotations, then equivalence does not hold for a set of states with nonzero
measure.

\subsection{Unequal spin}
The approach we use is to show that there is a state such that the strict
inequality
\begin{equation}
\label{strict}
V(J_x^{(-)})+V(J_y^{(+)}) > \langle J_z^{(+)} \rangle,
\end{equation}
holds under any combination of local unitary operations. From the continuity
of fidelity and the expectation values, there must be a finite region of states
near this state that are also not TMSS under any local unitaries. Hence, finding
a state satisfying \eqref{strict} is sufficient to show that the equivalence
between entanglement and two-mode spin squeezing fails to hold for a region of
states with nonzero measure.

For unequal spin, we consider the state for $j_1=\tfrac{1}{2}$ and $j_2=1$:
\begin{equation}
\label{count1}
\ket{\psi}=\frac{1}{\sqrt{2}}(\ket{\tfrac{1}{2} ,1}_z+\ket{-\tfrac{1}{2} ,0}_z).
\end{equation}
Because $[J_x^{(-)},J_y^{(+)}]=iJ_z^{(-)}$, the variances for $J_x^{(-)}$ and
$J_y^{(+)}$ satisfy
\begin{equation}
\label{genun}
V(J_x^{(-)})+V(J_y^{(+)}) \ge |\langle J_z^{(-)} \rangle|.
\end{equation}
{From} the derivation of this inequality, for equality it
would be necessary that $V(J_x^{(-)})=V(J_y^{(+)})=|\langle J_z^{(-)}
\rangle|/2$. The inequality $V(J_x^{(-)})V(J_y^{(+)})\ge|\langle J_z^{(-)}
\rangle|^2/4$ follows from the generalised uncertainty principle, and it is
known that equality is only possible if $(J_x^{(-)}-\lambda J_y^{(+)})
\ket{\phi}$ for some value of $\lambda$. If, in addition, $V(J_x^{(-)})=
V(J_y^{(+)})$, then it would follow that $(J_x^{(-)}-J_y^{(+)})\ket{\phi}=0$.
However, for $j_1=\tfrac{1}{2}$ and $j_2=1$, the determinant of $(J_x^{(-)}-
J_y^{(+)})$ is nonzero, so there is no state such that $(J_x^{(-)}-J_y^{(+)})
\ket{\phi}=0$. Hence it is not possible for equality to be obtained in
Eq.~\eqref{genun}.

For the specific example of the state $\ket\psi$ given in Eq.~\eqref{count1},
the reduced density matrix for subsystem 1 is the identity, so
$|\langle J_z^{(1)} \rangle|=0$ for $\ket\psi$ and all states related to
$\ket\psi$ by local unitaries. Thus $|\langle J_z^{(-)} \rangle|=
|\langle J_z^{(+)} \rangle|$, and we obtain the strict inequality
\eqref{strict}. Hence equivalence between TMSS and entanglement fails for a set
of states whose measure is not zero.

\subsection{Mixed states}
For the case of mixed states, we consider the example of the Werner state
\begin{equation}
\rho_\alpha = \alpha \ket{\Phi}\bra{\Phi} + \frac{1-\alpha}{(2J+1)^2} \openone ,
\end{equation}
with $\ket{\Phi}$ a maximally entangled state. This state is entangled for
$\alpha>1/(2J+2)$~\cite{horo}. In addition, the local reduced density matrices
are proportional to the identity, and therefore $\langle J_z^{(+)} \rangle$ must
be equal to zero under any local operations. In addition, because this state is
not maximally entangled, it can not satisfy $V(J_x^{(-)})=V(J_y^{(+)})=0$ (see
Appendix~B). Thus, the strict inequality \eqref{strict} is satisfied, and there
is a set of nonzero measure such that the equivalence between TMSS and
entanglement fails.

\subsection{Local rotations}
We can use a similar method for the case where the local operations are
restricted to rotations. For $j=1$, consider the state
\begin{equation}
	\ket{\psi}=(\ket{1,1}_z+\ket{-1,-1}_z)/\sqrt{2},
\end{equation}
which satisfies $\langle J_k^{(1,2)} \rangle = 0$, where $k\in\{x,y,z\}$.
Therefore, it will be the case that $\langle J_z^{(+)} \rangle=0$ under any
combination of local rotations. In addition, this state is not maximally
entangled, and therefore can not satisfy $V(J_x^{(-)})=V(J_y^{(+)})=0$. Thus,
this state must satisfy the strict inequality \eqref{strict} under any
combination of rotations. This means that, in addition to the state $\ket\psi$,
there is a finite region of states that are not equivalent to TMSS states under
local rotations.

\section{Conclusions}
We have shown that any entangled pure bipartite state of two subsystems with
equal spin is equivalent to a two-mode spin-squeezed state under local unitary
operations except for a set of states with measure zero. This equivalence
between two-mode spin squeezing and entanglement establishes a profound
connection between correlations that exceed the standard quantum limit, which is
so important for quantum metrology, especially high-resolution spectroscopy, and
entanglement, which provides a resource for quantum information processing.
Aside from the fundamental importance of this equivalence principle, a practical
merit of this result is that two-mode spin squeezing provides a macroscopic
signature of underlying entanglement, which will be useful in identifying the
presence of entanglement in physical systems.

Exceptions to this equivalence principle are maximally entangled states,
either in the full Hilbert space or in a restricted Hilbert space. Such states
comprise a set of measure zero, so the equivalence principle holds in an
approximate sense. We have shown, by providing counterexamples, that this
equivalence principle cannot be extended to the general cases of hybrid-spin
systems or mixed states.

\appendix
\section{}
In order to establish TMSS, we must evaluate the three expectation values in
Eq.\ \eqref{eq:three} for the state~(\ref{eq:psi}).
\begin{align}
\langle (J_x^{(1)})^2 \rangle &= \sum_{m_1,m_2=-j}^j \psi_{m_1}\psi_{m_2}\,
{_z\!\bra{m_1,m_1}} (J_x^{(1)})^2 \ket{m_2,m_2}_z 
= \sum_{m=-j}^j \psi_m^2\, {_z\!\bra{m}} (J_x^{(1)})^2 \ket{m}_z \nn \\
&= \sum_{m=-j}^j \psi_m^2\, {_z\!\bra{m}} J_x^{(1)} \big[ \alpha_{m-1}
\ket{m-1}_z+\alpha_m\ket{m+1}_z \big],
\end{align}
with $\alpha_m=\sqrt{j(j+1)-m(m+1)}/2$. We allow the states $\ket{-j-1}$ and
$\ket{j+1}$, provided they are multiplied by zero, which is the case here
because $\alpha_{-j-1}=\alpha_j=0$. This expression simplifies to
\begin{equation}
\label{one}
\langle (J_x^{(1)})^2 \rangle = \sum_{m=-j}^j \psi_m^2 (\alpha_{m-1}^2+
\alpha_m^2) = \frac 12 \sum_{m=-j}^j \psi_m^2 [j(j+1)-m^2].
\end{equation}
Similarly,
\begin{align}
\label{two}
\langle J_x^{(1)}J_x^{(2)} \rangle 
&= \sum_{m_1,m_2=-j}^j \psi_{m_1}\psi_{m_2}\, {_z\!\bra{m_1,
m_1}}\big[\alpha_{m_2-1}\ket{m_2-1}_z
+\alpha_{m_2}\ket{m_2+1}_z \big]\big[ \alpha_{m_2-1}\ket{m_2-1}_z
+\alpha_{m_2}|m_2+1\rangle_z \big]\nn \\
&= \sum_{m_1,m_2=-j}^j \psi_{m_1}\psi_{m_2}\, {_z\!\bra{m_1,m_1}}
\big[ \alpha_{m_2-1}^2 \ket{m_2-1,m_2-1}_z
+ \alpha_{m_2}^2 \ket{m_2+1,m_2+1}_z\big] \nn \\
&= \sum_{m=1-j}^j \psi_{m-1}\psi_m \alpha_{m-1}^2 
+ \sum_{m=-j}^{j-1} \psi_{m+1}\psi_m \alpha_m^2
= 2\sum_{m=-j}^{j-1} \psi_{m+1}\psi_m \alpha_m^2.
\end{align}
Lastly, evaluating $\langle J_z^{(+)}/2 \rangle$ gives
\begin{equation}
\label{three}
\langle J_z^{(+)}/2 \rangle = \sum_{m_1,m_2=-j}^j \psi_{m_1}\psi_{m_2}
\,{_z\!\bra{m_1,m_1}} m_2 \ket{m_2,m_2}_z
= \sum_{m=-j}^j m \psi_m^2.
\end{equation}

\section{}
Consider a state that satisfies
\begin{equation}
\label{eq:gencrit}
	V(J_y^{(+)}) = V(J_x^{(-)}) = 0 .
\end{equation}
This state must be an eigenstate of both $J_y^{(+)}$ and $J_x^{(-)}$. In
addition
\begin{equation}
\langle (J_y^{(+)})^n \rangle = \langle e^{i\frac{\pi}2 J_x^{(-)}}(J_y^{(+)})^n
e^{-i\frac{\pi}2 J_x^{(-)}} \rangle = \langle (J_z^{(-)})^n\rangle,
\end{equation}
where $n=1$ or 2. Therefore $V(J_z^{(-)})=0$, so the state is also an eigenstate
of $J_z^{(-)}$. Similarly we find
\begin{align}
\langle J_x^{(-)} \rangle &= \langle e^{-i\pi J_y^{(+)}} J_x^{(-)}
e^{i\pi J_y^{(+)}} \rangle = -\langle J_x^{(-)} \rangle, \nn \\
\langle J_y^{(+)} \rangle &= \langle e^{-i\pi J_x^{(-)}} J_y^{(+)}
e^{i\pi J_x^{(-)}} \rangle = -\langle J_y^{(+)} \rangle, \nn \\
\langle J_z^{(-)} \rangle &= \langle e^{-i\pi J_y^{(+)}} J_z^{(-)}
e^{i\pi J_y^{(+)}} \rangle = -\langle J_z^{(-)} \rangle .
\end{align}
Thus the state must be an eigenstate with eigenvalue zero of each of
$J_x^{(-)}$, $J_y^{(+)}$ and $J_z^{(-)}$. If the state is pure, it may be
written in any of the three forms
\begin{equation}
\label{forms}
\ket{\psi} = \sum_{m=-j_\text{min}}^{j_\text{min}} \psi_m \ket{m,m}_x
= \sum_{m=-j_\text{min}}^{j_\text{min}} \psi'_m\ket{m,-m}_y 
= \sum_{m=-j_\text{min}}^{j_\text{min}} \psi''_m \ket{m,m}_z .
\end{equation}
For generality we have allowed the possibility of unequal spins in the two
subsystems, and $j_\text{min}=\min\{j_1,j_2\}$. Note that the spins must be both
integer or both half-odd integer, because otherwise $J_x^{(-)}$, $J_y^{(+)}$
and $J_z^{(-)}$ could not have zero eigenvalues.

The reduced density matrices for subsystems~1 and~2, $\rho_1$ and~$\rho_2$,
commute with each of the local spin operators $J_x$, $J_y$, and~$J_z$, and
therefore are multiples of the identity. This implies that the spins for the
two subsystems are equal, and~$\ket{\psi}$ is maximally entangled.

We may show the corresponding result for a general mixed state $\rho$ from the
fact that the variance is a concave function of the state. A general mixed state
$\rho$ can be written in the form $\rho=\sum_k p_k \ket{\psi_k}\bra{\psi_k}$. If
$\rho$ satisfies Eq.\ \eqref{eq:gencrit}, then the individual pure states
$\ket{\psi_k}$ must satisfy Eq.\ \eqref{eq:gencrit} also. However, as shown
above the only pure state that satisfies Eq.\ \eqref{eq:gencrit} is the
maximally entangled state. Therefore this is the only solution if we allow mixed
states also.

\acknowledgments
This research has been supported by iCORE and the Australian Research Council.
DWB is grateful for valuable discussions with Robert W.\ Spekkens and Ernesto
F.\ Galv\~ao.

\end{document}